\documentclass[manuscript,screen,sigconf]{acmart}
\AtBeginDocument{
  }

\usepackage{listings}
\usepackage{todonotes}
\usepackage{xcolor}
\usepackage{subcaption}
\definecolor{nomicgreen}{HTML}{436444}
\definecolor{matryoshkared}{HTML}{D02907}
\definecolor{recombeegreen}{HTML}{3BC4A1}
\definecolor{reconstructedpink}{HTML}{BF1778}

\lstdefinestyle{mystyle}{
  language=Python,
  basicstyle=\ttfamily\small,
  keywordstyle=\color{blue},
  commentstyle=\color{gray},
  stringstyle=\color{orange},
  showstringspaces=false,
  columns=flexible,
  keepspaces=true,
  frame=single,
  breaklines=true,
  breakatwhitespace=true,
  tabsize=2,
  captionpos=b
}
\setcopyright{rightsretained}
\copyrightyear{2025}
\acmYear{2025}
\acmDOI{XXXXXXX.XXXXXXX}
\acmConference[Under review]{}{2025}{Prague, Czech Republic}
\acmBooktitle{X}
\acmISBN{X}

\renewcommand\footnotetextcopyrightpermission[1]{}
\begin{document}

\title{The Future is Sparse: Embedding Compression for Scalable Retrieval in Recommender Systems}

\author{Petr Kasalický$^{1,2}$, Martin Spišák$^{1,3}$, Vojtěch Vančura$^{1,2}$\\Daniel Bohuněk$^{1}$, Rodrigo Alves$^{1,2}$, Pavel Kordík$^{1,2}$}
\affiliation{
  \institution{$^1$Recombee, $^2$Czech Technical University in Prague, $^3$Charles University}
  \city{Prague}
  \country{Czech Republic}
}
\email{{petr.kasalicky,martin.spisak,vojtech.vancura,daniel.bohunek,rodrigo.alves,pavel.kordik}@recombee.com}

\renewcommand{\shortauthors}{Recombee}

\begin{abstract}
Industry-scale recommender systems
face a core challenge: representing entities with high cardinality, such as users or items, using dense embeddings that must be accessible during both training and inference. However, as embedding sizes grow, memory constraints make storage and access increasingly 
difficult.
We describe a lightweight, learnable embedding compression technique that projects dense embeddings into a high-dimensional, sparsely activated space. Designed for retrieval tasks, our method reduces memory requirements while preserving retrieval performance, enabling scalable deployment under strict resource constraints. Our results demonstrate that \emph{leveraging sparsity} is a promising approach for improving the efficiency of large-scale recommenders. We release our code at \url{https://github.com/recombee/CompresSAE}.
\end{abstract}


\maketitle

\section{Introduction}

Industrial recommender systems (RSs) pursue extreme performance at scale, handling massive embedding tables of dense, high dimensional user and item vectors spanning multiple modalities: textual, visual, video, and behavioral~\cite{liu2022monolith}. These expressive representations are essential for modeling fine-grained user preferences and item characteristics. However, as embedding dimensions are progressively increasing, so do the costs:  memory, flops, and retrieval latency all grow sharply, challenging large-scale deployments.

This paper presents our experience with improving the quality-efficiency trade-off in a production setting at \textbf{Recombee}\footnote{\url{https://www.recombee.com/}}, an AI-powered Recommender-as-a-Service platform. We focus on at-scale deployments, managing catalogs with $\mathcal{O}(10^8)$ richly annotated items and serving high-throughput personalized traffic. As in many real-world systems, these scenarios routinely suffer from persistent cold-start challenges, with a significant fraction of the catalog never receiving interactions. This long-tail sparsity motivates the use of powerful content-based embeddings to enable catalog exploration. However, doing so further exacerbates the efficiency crisis by inflating embedding tables.

\begin{figure}[t]
    \centering
    \begin{minipage}[t]{\linewidth}
        \centering
        \setlength{\tabcolsep}{4pt}
        \begin{tabular}{@{}lrrr@{}}
            \toprule
            \multicolumn{1}{l}{Model} & \multicolumn{1}{r}{Embedding} & \multicolumn{1}{r}{CTR} & \multicolumn{1}{r}{Size per 100M} \\
            \multicolumn{1}{l}{(Compression)} & \multicolumn{1}{r}{Dimension} & \multicolumn{1}{r}{Lift} & \multicolumn{1}{r}{Embeddings} \\
            \midrule
            SBERT~\cite{reimers-2019-sentence-bert} & 512 & (baseline) & 204.8 GB\\
            \textcolor{nomicgreen}{Nomic}~\cite{nussbaum2024nomic} & 768 & $+4.86\%$ & 307.2 GB\\
            \midrule
            \textcolor{matryoshkared}{Nomic} (Matryoshka) & 64 & $+1.89\%$ & 25.6 GB\\
            \textcolor{recombeegreen}{Nomic} (C\textsc{ompres}SAE) & 4096* & \textbf{+3.44\%} & 25.6 GB\\
            \bottomrule
        \end{tabular}
        \begin{flushleft}
        \footnotesize{*Sparse embeddings with 32 nonzero entries.}
        \end{flushleft}
    \end{minipage}
    \begin{minipage}[t]{\linewidth}
        \centering
        \includegraphics[width=\linewidth]{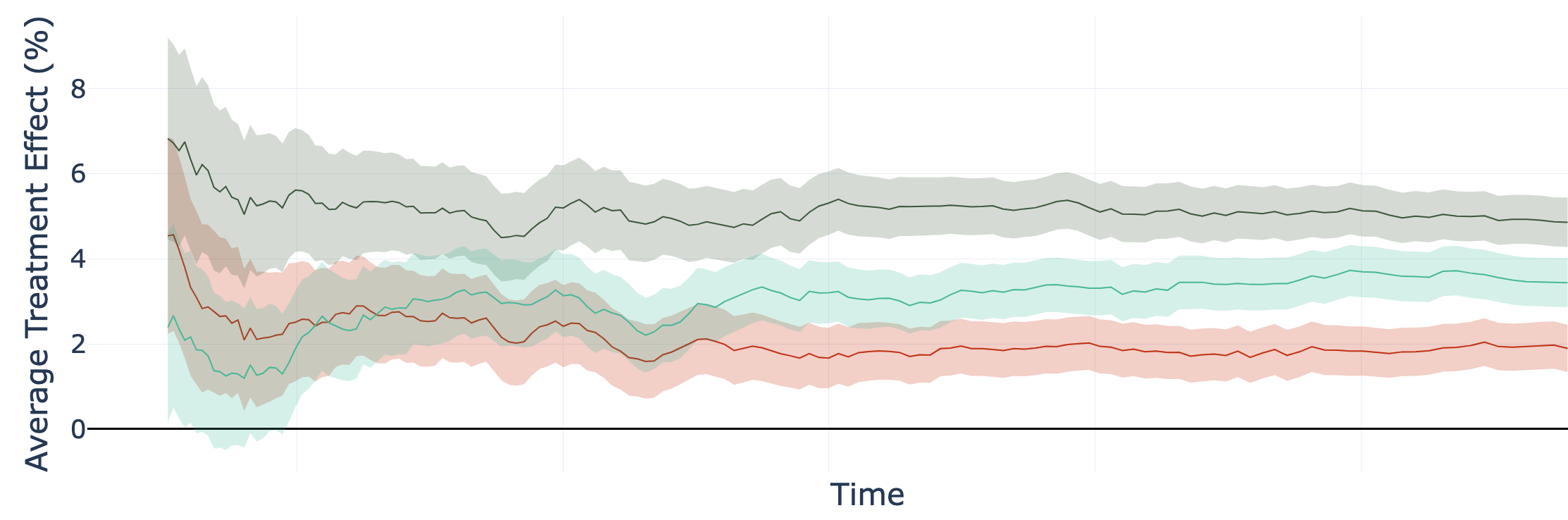}
        \caption{Comparison of embedding models used for candidate retrieval. We report online recommendation performance on a downstream task, relative to SBERT~\cite{reimers-2019-sentence-bert}, with anytime-valid 99\% confidence intervals.}
        \label{fig1}
    \end{minipage}
\end{figure}

As is standard in industry, Recombee systems employ multistage architectures~\cite{10.1145/2959100.2959190} to manage complexity and performance, with expressive embeddings serving downstream tasks like retrieval, ranking, clustering, and personalization. In the candidate retrieval stage, in particular, high-quality embeddings can measurably improve candidate set recall. For example, replacing the production SBERT (\texttt{distil-use-base-multilingual-cased-v2})~\cite{reimers-2019-sentence-bert} with a Nomic model (\texttt{nomic-embed-text-v1.5})~\cite{nussbaum2024nomic} recently yielded a 4.86\% uplift in click-through rate (CTR), see Figure~\ref{fig1}. But such gains come at a steep price: for catalogs with 100 million items, embedding tables routinely reach hundreds of GBs, placing pressure on storage, distribution, caching, and accelerator memory, and ultimately limiting both training and inference throughput.

The industry trend is clear: larger embeddings improve quality. But they are also slower, more costly, and less scalable. While recent approaches like Matryoshka~\cite{kusupati2024matryoshkarepresentationlearning} compress embeddings via progressive truncation and are gaining traction, they come at the cost of retraining the backbone model. In contrast, we advocate for a different approach: \textbf{sparse embeddings}. By maintaining high dimensionality but enforcing structured sparsity, we retain representational capacity while dramatically reducing memory and compute overhead. This direction offers a scalable and efficient path forward for retrieval in next-generation recommender systems.

\begin{figure*}[ht!]
    \centering
    \includegraphics[width=\textwidth]{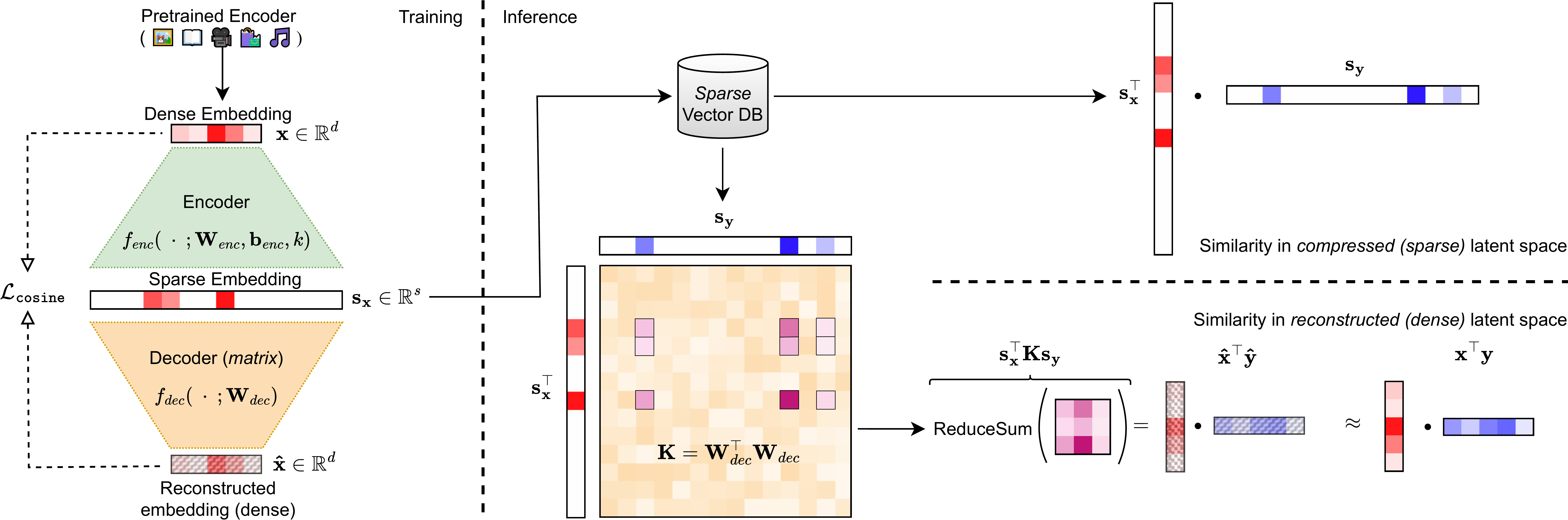}
    \caption{Left: Training. Right: Two modes of inference}
    \label{fig2}
    \vspace*{-0.2cm}
\end{figure*}

\section{Related Work}
Numerous strategies have been proposed to address the challenges related to large embedding tables in RSs, each with distinct trade-offs~\cite{li2024embedding_survey}. For instance, \emph{quantization} (e.g., \texttt{int4}~\cite{post_training_4bit}) offers a strong compression–accuracy trade-off and is widely adopted, though it requires hardware-specific support and quantization-aware retraining~\cite{li2023adaptive_low_precision}. \emph{Dimensionality reduction} via SVD or PCA is efficient, but typically incurs a notable accuracy loss~\cite{raunak2019dim_reduction}. In contrast, \emph{Matryoshka representation learning}~\cite{kusupati2024matryoshkarepresentationlearning} enables effective truncation by training nested encoders, though at the cost of retraining the backbone. \emph{Offloading and caching} techniques, such as Meta’s multi-tier memory systems~\cite{10.1145/3640457.3688037}, scale to 100 TB tables but requires complex engineering and platform dependence. In practice, these methods are often combined to achieve effective compression at scale.

Recently, \emph{sparse autoencoders} (SAEs) have emerged as a compelling alternative. Rooted in self-supervised dictionary learning~\cite{makhzani2014ksparseautoencoders}, SAEs have gained traction for producing interpretable and disentangled activations~\cite{bricken2023towards,cunningham2023sparseautoencodershighlyinterpretable}, including in frontier LLMs~\cite{templeton2024scaling,gao2024scalingevaluatingsparseautoencoders}. Recent work~\cite{wen2025matryoshkarevisitingsparsecoding} extends the architecture of~\cite{gao2024scalingevaluatingsparseautoencoders} by incorporating a contrastive loss on the sparse latent representations to further disentangle them, achieving compression quality superior to Matryoshka of equal size across several offline experiments. Our method follows this paradigm, being able to efficiently handle tens to hundreds of millions of items.

\section{Approach}
We start by training an SAE on a corpus of $N$ $d$-dimensional embeddings $\mathbf{E} \in \mathbb{R}^{N \times d}$ produced by a pretrained encoder (similarly to~\cite{wen2025matryoshkarevisitingsparsecoding}) -- i.e., without finetuning the encoder itself -- making our approach more flexible and computationally cheaper compared to Matryoshka-based compression. However, unlike \cite{wen2025matryoshkarevisitingsparsecoding}, which extends the interpretability-focused architecture of~\cite{gao2024scalingevaluatingsparseautoencoders} by adding a contrastive loss, we employ a new architecture specifically designed for the use of the embeddings in retrieval tasks, with a primary emphasis on preserving directional information. 

Our sparse autoencoder, C\textsc{ompres}SAE, comprises a non-linear encoder $f_{enc}$, with learnable weights $\mathbf{W}_{enc} \in \mathbb{R}^{h \times d}$, bias $\mathbf{b}_{enc} \in \mathbb{R}^{h}$, and a positive integer sparsity hyperparameter $k$, that encodes an input vector $\mathbf{x} \in \mathbb{R}^{d}$ into a $k$-sparse $h$-dimensional latent representation $\mathbf{s} \in \mathbb{R}^{h}$. This sparse representation is then reconstructed using a \emph{linear} decoder $f_{dec}$, parametrized by $\mathbf{W}_{dec} \in \mathbb{R}^{d \times h}$. Written formally:
\begin{align}
   \mathbf{s} & = f_{enc}(\mathbf{x};\mathbf{W}_{enc},\mathbf{b}_{enc},k) = \phi(\mathbf{W}_{enc}\mathbf{\overline{x}} + \mathbf{b}_{enc},k)\\
   \mathbf{\hat{x}} & = f_{dec}(\mathbf{s}; \mathbf{W}_{dec}) = \mathbf{W}_{dec}\mathbf{s}
\end{align}
Different from prior works~\cite{wen2025matryoshkarevisitingsparsecoding,gao2024scalingevaluatingsparseautoencoders} which standardize the input $\mathbf{x}$ before encoding and later rescale the output, we simply normalize the input: $\mathbf{\overline{x}} = \mathbf{x}/\|\mathbf{x}\|_2$. Note that this is directly aligned with retrieval use cases, where cosine similarity is used. Moreover, $\phi(\cdot,k): \mathbb{R}^h \rightarrow \mathbb{R}^h$, a function that retains the $k$ entries with the largest \emph{absolute} values and zeroes out the rest, serves both as a sparsification mechanism and non-linear activation in our model, replacing the common choice of ReLU and TopK. This design promotes the distillation of the original vector's direction into the sparse latent space and better utilizes its limited capacity by also preserving negative values. Finally, the decoder is bias-free, which enables a retrieval "trick" described in Section~\ref{sec32}, and its weight matrix $\mathbf{W}_{\text{dec}}$ is row-normalized, ensuring consistent output scaling.

Another innovative aspect of our SAE from prior approaches is the choice of reconstruction objective. Our goal is to use the embeddings for similarity search with cosine distance as the metric; thus, we depart from the traditional $\ell_2$-reconstruction and instead train the model to minimize the \emph{cosine distance} between the input $\mathbf{x}$ and its reconstruction $\mathbf{\hat{x}} = f(\mathbf{x};\theta,k)$:
\begin{equation}
    \mathcal{L}_{\texttt{cosine}}\big(\mathbf{x},f(\mathbf{x};\theta,k)\big) = 1 - \frac{\mathbf{x}^\top\mathbf{\hat{x}}}{\|\mathbf{x}\|_2\|\mathbf{\hat{x}}\|_2}
\end{equation}
Here, $f = f_{dec} \circ f_{enc}$ and $\theta$ denotes the learnable parameters of the SAE.
Analogous to~\cite{gao2024scalingevaluatingsparseautoencoders}, our final loss $\mathcal{L}$ combines $\mathcal{L}_{\texttt{cosine}}$ for reconstructions with different $k$s, $f(\mathbf{x};\theta,k)$ and $f(\mathbf{x};\theta,4k)$, which has proven effective in the preventing emergence of dead neurons.

\begin{figure*}[ht]
  \centering
  \begin{subfigure}[b]{0.32\textwidth}
    \centering
    \includegraphics[width=\textwidth]{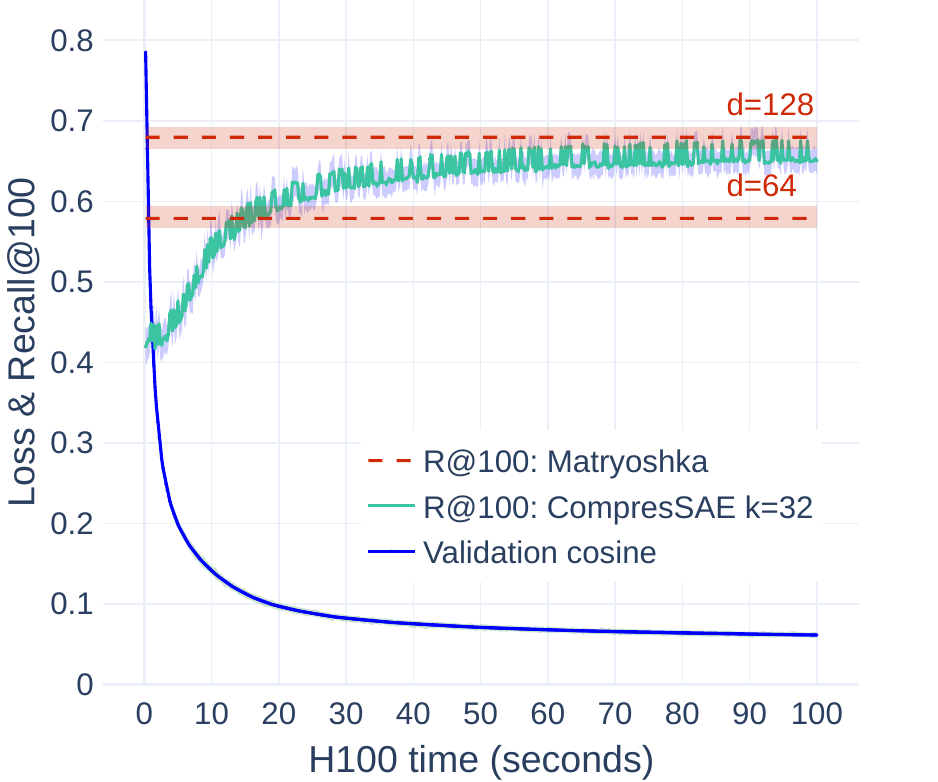}
  \end{subfigure}
  \hfill
  \begin{subfigure}[b]{0.32\textwidth}
    \centering
    \includegraphics[width=\textwidth]{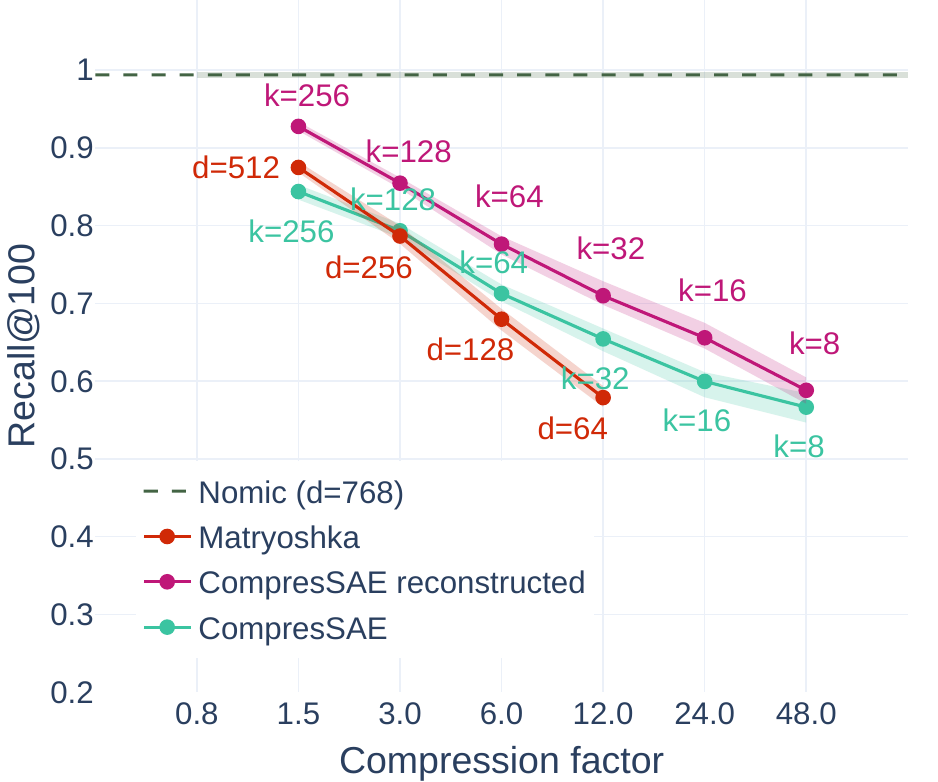}
  \end{subfigure}
  \hfill
  \begin{subfigure}[b]{0.32\textwidth}
    \centering
    \includegraphics[width=\textwidth]{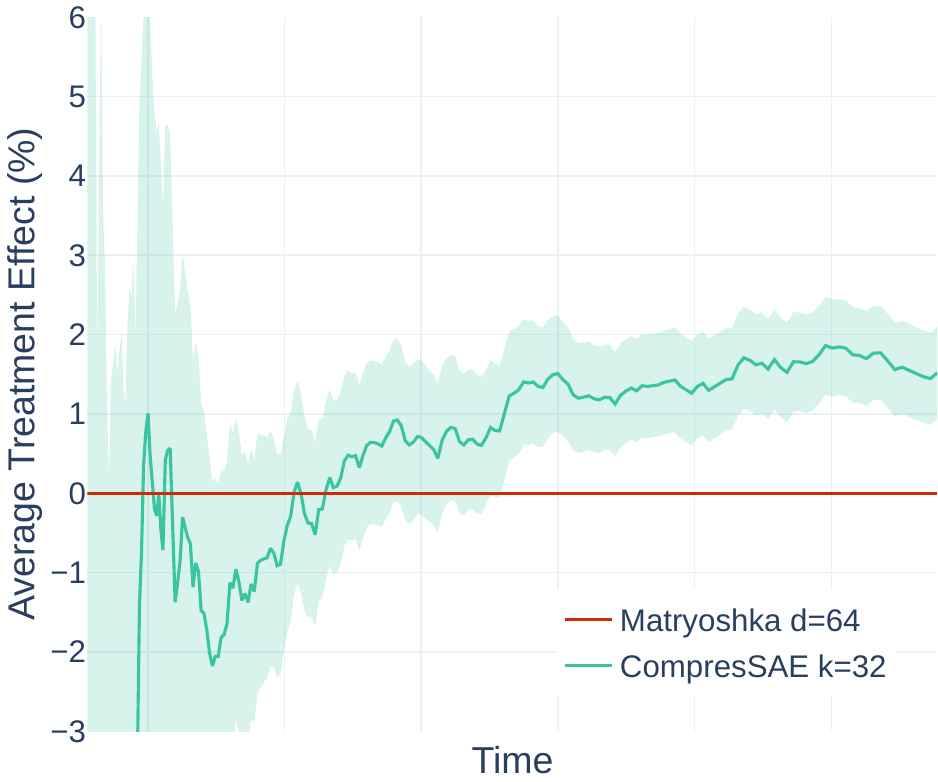}
  \end{subfigure}
  \caption{
    Left: Training convergence. 
    Center: Accuracy–compression trade-off. 
    Right: A/B test results (baseline: \textcolor{matryoshkared}{Matryoshka}).
  }
  \label{fig3}
\end{figure*}

\subsection{Training}
As shown in Figure \ref{fig2} (left), C\textsc{ompres}SAE operates directly on batches of precomputed embeddings, requiring no access to the original encoder or raw data. Its compact architecture supports large batch sizes, enabling efficient and fast training. Figure \ref{fig3} (left) shows a run using the Adam optimizer~\cite{kingma2017adammethodstochasticoptimization} with a batch size of 100~000, reaching convergence in approximately 500 steps (around 100 seconds) on a single NVIDIA H100 SXM 80GB GPU.

\subsection{Inference}\label{sec32}
\noindent\textbf{Sparse Embedding Compression.} To compress embeddings, we pass dense batches $\mathbf{E}_{\texttt{batch}} \in \mathbb{R}^{b \times d}$ through the trained C\textsc{ompres}SAE encoder $f_{enc}$ and store the resulting sparse outputs $\mathbf{S}_{\texttt{batch}} \in \mathbb{R}^{b \times s}$ in Compressed Sparse Row (CSR) format~\cite{https://doi.org/10.1002/nme.1620180804}, enabling efficient storage and dot-product computation by exploiting sparsity. Specifically, a \texttt{float32} sparse embedding with $k$ nonzero entries requires $2\cdot k \cdot 4$ bytes to store its values and indices. For example, compressing a 768-dimensional dense embedding into a 4096-dimensional sparse vector with 32 nonzeros yields an effective $12\times$ compression. 

In practice, CSR format is widely supported by popular frameworks (e.g., SciPy~\cite{2020SciPy-NMeth}, PyTorch~\cite{NEURIPS2019_9015}) and vector databases such as PostgresSQL's pgvector\footnote{\url{https://github.com/pgvector/pgvector}}.

\noindent\textbf{Retrieval from Sparse Compressed Space.}
The sparsity of our compressed embeddings also unlocks significant retrieval efficiency: dot products between vectors with $k$ nonzero entries have O($k$) complexity, \emph{independent of the embedding dimensionality}. This enables fast cosine similarity computation using sparse matrix–vector (SpMV) kernels, even in high-dimensional spaces. As with dense embeddings, retrieval consists of computing similarities between vectors, followed by a top-$n$ selection that can be performed either exactly or approximately using an index -- e.g., pgvector supports approximate search via HNSW~\cite{8594636}.

\noindent\textbf{Retrieval from \emph{Reconstructed} Space.}
The linear design of decoder $f_{dec}$ enables a second, slower but more precise retrieval method: computing similarity in the reconstructed space via a kernel trick. This space is accurately aligned with the original embedding space, as a consequence of our cosine reconstruction objective between inputs $\mathbf{x}$ and reconstructions $\mathbf{\hat{x}}$; formally:

\begin{equation*}
   \frac{\mathbf{x}^\top\mathbf{y}}{\|\mathbf{x}\|_2\|\mathbf{y}\|_2} 
   \approx \frac{\mathbf{\hat{x}}^\top\mathbf{\hat{y}}}{\|\mathbf{\hat{x}}\|_2\|\mathbf{\hat{y}}\|_2} 
   = \frac{(\mathbf{W}_{dec}\mathbf{s}_\mathbf{x})^\top(\mathbf{W}_{dec}\mathbf{s}_\mathbf{y})}{\|\mathbf{W}_{dec}\mathbf{s}_\mathbf{x}\|_2\|\mathbf{W}_{dec}\mathbf{s}_\mathbf{y}\|_2} 
   = \frac{\mathbf{s}_\mathbf{x}^\top\mathbf{K}\mathbf{s}_\mathbf{y}}{\sqrt{\mathbf{s}_\mathbf{x}^\top\mathbf{K}\mathbf{s}_\mathbf{x}} \sqrt{\mathbf{s}_\mathbf{y}^\top\mathbf{K}\mathbf{s}_\mathbf{y}}} 
\end{equation*}

Here, similarity is computed using only the sparse representations $\mathbf{s}_\mathbf{x}$ and $\mathbf{s}_\mathbf{y}$ and the kernel matrix $\mathbf{K} = \mathbf{W}_{dec}^\top\mathbf{W}_{dec} \in \mathbb{R}^{s\times s}$. Since both $\mathbf{s}_\mathbf{x}$ and $\mathbf{s}_\mathbf{y}$ contain only $k$ nonzero entries, the complexity of this operation is $O(k^2)$, which remains efficient when $k$ is small. This retrieval path enables high-fidelity similarity estimation from sparse embeddings, with a modest increase in computational cost.

\section{Experimental Results}
We conducted our experiments with a proprietary dataset from a global client in the media domain, with a catalog of $\mathcal{O}(10^8)$ items.

\noindent\textbf{Offline Experiments.} Figure~\ref{fig3} (left) demonstrates the fast training convergence of C\textsc{ompres}SAE, surpassing the retrieval accuracy of equally sized Matryoshka after approximately 15 seconds. The accuracy-compression trade-off is compared in Figure~\ref{fig3} (center). C\textsc{ompres}SAE achieves superior balance between compression and accuracy compared to Matryoshka, especially at high compression ratios: \textcolor{recombeegreen}{C\textsc{ompres}SAE} can match the performance of Matryoshka models up to four times larger. Finally, \textcolor{reconstructedpink}{retrieval from the reconstructed space} yields the best trade-off overall.

\noindent\textbf{Online Experiment} (Figure~\ref{fig1}, Figure~\ref{fig3} (right)). We also conducted an A/B test for online evaluation. Each variant received a sample of approximately 8.5 million users. The experiment evaluates four different embedding models used for candidate retrieval: 512-dimensional SBERT (baseline); 768-dimensional \textcolor{nomicgreen}{Nomic} model trained with Matryoshka loss; its 64-dimensional \textcolor{matryoshkared}{Matryoshka} variant; and \textcolor{recombeegreen}{C\textsc{ompres}SAE}, our 4096-dimensional sparse compression of the Nomic model with $k=32$, where similarity is computed directly in the sparse compressed space. Performance is measured via downstream CTR. 

Results show that both Matryoshka embeddings and C\textsc{ompres}SAE outperform the $8\times$ larger SBERT (Figure~\ref{fig1}). Compared to the uncompressed Nomic model, our $12\times$ compressed C\textsc{ompres}SAE sacrifices only 1.35\% of CTR performance. Notably, our lightweight post-processing method statistically significantly outperforms Matryoshka of the same size by $+1.52\%$ (Figure~\ref{fig3}, right).

\section{Conclusions}

We introduce C\textsc{ompres}SAE, a sparse autoencoder that compresses dense embeddings into high-dimensional, sparsely activated vectors optimized for candidate retrieval. The method preserves cosine similarity, requires no retraining of the original encoder, and delivers significant memory and compute savings without sacrificing downstream performance.
Online deployment at Recombee confirms that C\textsc{ompres}SAE outperforms leading Matryoshka-based compression in retrieval quality, as measured by downstream CTR.

\emph{Our results establish sparsity as a promising direction for enhancing the efficiency of web-scale recommender systems.}

\bibliographystyle{ACM-Reference-Format}
\bibliography{references}


\begin{thebibliography}{21}


\ifx \showCODEN    \undefined \def \showCODEN     #1{\unskip}     \fi
\ifx \showISBNx    \undefined \def \showISBNx     #1{\unskip}     \fi
\ifx \showISBNxiii \undefined \def \showISBNxiii  #1{\unskip}     \fi
\ifx \showISSN     \undefined \def \showISSN      #1{\unskip}     \fi
\ifx \showLCCN     \undefined \def \showLCCN      #1{\unskip}     \fi
\ifx \shownote     \undefined \def \shownote      #1{#1}          \fi
\ifx \showarticletitle \undefined \def \showarticletitle #1{#1}   \fi
\ifx \showURL      \undefined \def \showURL       {\relax}        \fi
\providecommand\bibfield[2]{#2}
\providecommand\bibinfo[2]{#2}
\providecommand\natexlab[1]{#1}
\providecommand\showeprint[2][]{arXiv:#2}

\bibitem[Bricken et~al\mbox{.}(2023)]%
        {bricken2023towards}
\bibfield{author}{\bibinfo{person}{Trenton Bricken}, \bibinfo{person}{Adly Templeton}, \bibinfo{person}{Joshua Batson}, \bibinfo{person}{Brian Chen}, \bibinfo{person}{Adam Jermyn}, \bibinfo{person}{Tom Conerly}, \bibinfo{person}{Nick Turner}, \bibinfo{person}{Cem Anil}, \bibinfo{person}{Carson Denison}, \bibinfo{person}{Amanda Askell}, {et~al\mbox{.}}} \bibinfo{year}{2023}\natexlab{}.
\newblock \showarticletitle{Towards monosemanticity: Decomposing language models with dictionary learning}.
\newblock \bibinfo{journal}{\emph{Transformer Circuits Thread}}  \bibinfo{volume}{2} (\bibinfo{year}{2023}).
\newblock


\bibitem[Covington et~al\mbox{.}(2016)]%
        {10.1145/2959100.2959190}
\bibfield{author}{\bibinfo{person}{Paul Covington}, \bibinfo{person}{Jay Adams}, {and} \bibinfo{person}{Emre Sargin}.} \bibinfo{year}{2016}\natexlab{}.
\newblock \showarticletitle{Deep Neural Networks for YouTube Recommendations}. In \bibinfo{booktitle}{\emph{Proceedings of the 10th ACM Conference on Recommender Systems}} (Boston, Massachusetts, USA) \emph{(\bibinfo{series}{RecSys '16})}. \bibinfo{publisher}{Association for Computing Machinery}, \bibinfo{address}{New York, NY, USA}, \bibinfo{pages}{191–198}.
\newblock
\showISBNx{9781450340359}
\href{https://doi.org/10.1145/2959100.2959190}{doi:\nolinkurl{10.1145/2959100.2959190}}


\bibitem[Cunningham et~al\mbox{.}(2023)]%
        {cunningham2023sparseautoencodershighlyinterpretable}
\bibfield{author}{\bibinfo{person}{Hoagy Cunningham}, \bibinfo{person}{Aidan Ewart}, \bibinfo{person}{Logan Riggs}, \bibinfo{person}{Robert Huben}, {and} \bibinfo{person}{Lee Sharkey}.} \bibinfo{year}{2023}\natexlab{}.
\newblock \bibinfo{title}{Sparse Autoencoders Find Highly Interpretable Features in Language Models}.
\newblock
\showeprint[arxiv]{2309.08600}~[cs.LG]
\urldef\tempurl%
\url{https://arxiv.org/abs/2309.08600}
\showURL{%
\tempurl}


\bibitem[Eisenstat et~al\mbox{.}(1982)]%
        {https://doi.org/10.1002/nme.1620180804}
\bibfield{author}{\bibinfo{person}{S.~C. Eisenstat}, \bibinfo{person}{M.~C. Gursky}, \bibinfo{person}{M.~H. Schultz}, {and} \bibinfo{person}{A.~H. Sherman}.} \bibinfo{year}{1982}\natexlab{}.
\newblock \showarticletitle{Yale sparse matrix package I: The symmetric codes}.
\newblock \bibinfo{journal}{\emph{Internat. J. Numer. Methods Engrg.}} \bibinfo{volume}{18}, \bibinfo{number}{8} (\bibinfo{year}{1982}), \bibinfo{pages}{1145--1151}.
\newblock
\href{https://doi.org/10.1002/nme.1620180804}{doi:\nolinkurl{10.1002/nme.1620180804}}
\showeprint{https://onlinelibrary.wiley.com/doi/pdf/10.1002/nme.1620180804}


\bibitem[Gao et~al\mbox{.}(2024)]%
        {gao2024scalingevaluatingsparseautoencoders}
\bibfield{author}{\bibinfo{person}{Leo Gao}, \bibinfo{person}{Tom~Dupré la Tour}, \bibinfo{person}{Henk Tillman}, \bibinfo{person}{Gabriel Goh}, \bibinfo{person}{Rajan Troll}, \bibinfo{person}{Alec Radford}, \bibinfo{person}{Ilya Sutskever}, \bibinfo{person}{Jan Leike}, {and} \bibinfo{person}{Jeffrey Wu}.} \bibinfo{year}{2024}\natexlab{}.
\newblock \bibinfo{title}{Scaling and evaluating sparse autoencoders}.
\newblock
\showeprint[arxiv]{2406.04093}~[cs.LG]
\urldef\tempurl%
\url{https://arxiv.org/abs/2406.04093}
\showURL{%
\tempurl}


\bibitem[Guan et~al\mbox{.}(2019)]%
        {post_training_4bit}
\bibfield{author}{\bibinfo{person}{Hui Guan}, \bibinfo{person}{Andrey Malevich}, \bibinfo{person}{Jiyan Yang}, \bibinfo{person}{Jongsoo Park}, {and} \bibinfo{person}{Hector Yuen}.} \bibinfo{year}{2019}\natexlab{}.
\newblock \showarticletitle{Post-Training 4-bit Quantization on Embedding Tables}.
\newblock \bibinfo{journal}{\emph{CoRR}}  \bibinfo{volume}{abs/1911.02079} (\bibinfo{year}{2019}).
\newblock
\urldef\tempurl%
\url{http://arxiv.org/abs/1911.02079}
\showURL{%
\tempurl}


\bibitem[Kingma and Ba(2017)]%
        {kingma2017adammethodstochasticoptimization}
\bibfield{author}{\bibinfo{person}{Diederik~P. Kingma} {and} \bibinfo{person}{Jimmy Ba}.} \bibinfo{year}{2017}\natexlab{}.
\newblock \bibinfo{title}{Adam: A Method for Stochastic Optimization}.
\newblock
\showeprint[arxiv]{1412.6980}~[cs.LG]
\urldef\tempurl%
\url{https://arxiv.org/abs/1412.6980}
\showURL{%
\tempurl}


\bibitem[Kusupati et~al\mbox{.}(2024)]%
        {kusupati2024matryoshkarepresentationlearning}
\bibfield{author}{\bibinfo{person}{Aditya Kusupati}, \bibinfo{person}{Gantavya Bhatt}, \bibinfo{person}{Aniket Rege}, \bibinfo{person}{Matthew Wallingford}, \bibinfo{person}{Aditya Sinha}, \bibinfo{person}{Vivek Ramanujan}, \bibinfo{person}{William Howard-Snyder}, \bibinfo{person}{Kaifeng Chen}, \bibinfo{person}{Sham Kakade}, \bibinfo{person}{Prateek Jain}, {and} \bibinfo{person}{Ali Farhadi}.} \bibinfo{year}{2024}\natexlab{}.
\newblock \bibinfo{title}{Matryoshka Representation Learning}.
\newblock
\showeprint[arxiv]{2205.13147}~[cs.LG]
\urldef\tempurl%
\url{https://arxiv.org/abs/2205.13147}
\showURL{%
\tempurl}


\bibitem[Li et~al\mbox{.}(2023)]%
        {li2023adaptive_low_precision}
\bibfield{author}{\bibinfo{person}{Shiwei Li}, \bibinfo{person}{Huifeng Guo}, \bibinfo{person}{Lu Hou}, \bibinfo{person}{Wei Zhang}, \bibinfo{person}{Xing Tang}, \bibinfo{person}{Ruiming Tang}, \bibinfo{person}{Rui Zhang}, {and} \bibinfo{person}{Ruixuan Li}.} \bibinfo{year}{2023}\natexlab{}.
\newblock \showarticletitle{Adaptive low-precision training for embeddings in click-through rate prediction}. In \bibinfo{booktitle}{\emph{Proceedings of the AAAI Conference on Artificial Intelligence}}, Vol.~\bibinfo{volume}{37}. \bibinfo{pages}{4435--4443}.
\newblock


\bibitem[Li et~al\mbox{.}(2024)]%
        {li2024embedding_survey}
\bibfield{author}{\bibinfo{person}{Shiwei Li}, \bibinfo{person}{Huifeng Guo}, \bibinfo{person}{Xing Tang}, \bibinfo{person}{Ruiming Tang}, \bibinfo{person}{Lu Hou}, \bibinfo{person}{Ruixuan Li}, {and} \bibinfo{person}{Rui Zhang}.} \bibinfo{year}{2024}\natexlab{}.
\newblock \showarticletitle{Embedding compression in recommender systems: A survey}.
\newblock \bibinfo{journal}{\emph{Comput. Surveys}} \bibinfo{volume}{56}, \bibinfo{number}{5} (\bibinfo{year}{2024}), \bibinfo{pages}{1--21}.
\newblock


\bibitem[Liu et~al\mbox{.}(2022)]%
        {liu2022monolith}
\bibfield{author}{\bibinfo{person}{Zhuoran Liu}, \bibinfo{person}{Leqi Zou}, \bibinfo{person}{Xuan Zou}, \bibinfo{person}{Caihua Wang}, \bibinfo{person}{Biao Zhang}, \bibinfo{person}{Da Tang}, \bibinfo{person}{Bolin Zhu}, \bibinfo{person}{Yijie Zhu}, \bibinfo{person}{Peng Wu}, \bibinfo{person}{Ke Wang}, {et~al\mbox{.}}} \bibinfo{year}{2022}\natexlab{}.
\newblock \showarticletitle{Monolith: real time recommendation system with collisionless embedding table}.
\newblock \bibinfo{journal}{\emph{arXiv preprint arXiv:2209.07663}} (\bibinfo{year}{2022}).
\newblock


\bibitem[Makhzani and Frey(2014)]%
        {makhzani2014ksparseautoencoders}
\bibfield{author}{\bibinfo{person}{Alireza Makhzani} {and} \bibinfo{person}{Brendan Frey}.} \bibinfo{year}{2014}\natexlab{}.
\newblock \bibinfo{title}{k-Sparse Autoencoders}.
\newblock
\showeprint[arxiv]{1312.5663}~[cs.LG]
\urldef\tempurl%
\url{https://arxiv.org/abs/1312.5663}
\showURL{%
\tempurl}


\bibitem[Malkov and Yashunin(2020)]%
        {8594636}
\bibfield{author}{\bibinfo{person}{Yu~A. Malkov} {and} \bibinfo{person}{D.~A. Yashunin}.} \bibinfo{year}{2020}\natexlab{}.
\newblock \showarticletitle{Efficient and Robust Approximate Nearest Neighbor Search Using Hierarchical Navigable Small World Graphs}.
\newblock \bibinfo{journal}{\emph{IEEE Transactions on Pattern Analysis and Machine Intelligence}} \bibinfo{volume}{42}, \bibinfo{number}{4} (\bibinfo{year}{2020}), \bibinfo{pages}{824--836}.
\newblock
\href{https://doi.org/10.1109/TPAMI.2018.2889473}{doi:\nolinkurl{10.1109/TPAMI.2018.2889473}}


\bibitem[Nussbaum et~al\mbox{.}(2024)]%
        {nussbaum2024nomic}
\bibfield{author}{\bibinfo{person}{Zach Nussbaum}, \bibinfo{person}{John~X. Morris}, \bibinfo{person}{Brandon Duderstadt}, {and} \bibinfo{person}{Andriy Mulyar}.} \bibinfo{year}{2024}\natexlab{}.
\newblock \bibinfo{title}{Nomic Embed: Training a Reproducible Long Context Text Embedder}.
\newblock
\showeprint[arxiv]{2402.01613}~[cs.CL]


\bibitem[Park et~al\mbox{.}(2024)]%
        {10.1145/3640457.3688037}
\bibfield{author}{\bibinfo{person}{Intaik Park}, \bibinfo{person}{Ehsan Ardestani}, \bibinfo{person}{Damian Reeves}, \bibinfo{person}{Sarunya Pumma}, \bibinfo{person}{Henry Tsang}, \bibinfo{person}{Levy Zhao}, \bibinfo{person}{Jian He}, \bibinfo{person}{Joshua Deng}, \bibinfo{person}{Dennis Van~der Staay}, \bibinfo{person}{Yu Guo}, {and} \bibinfo{person}{Paul Zhang}.} \bibinfo{year}{2024}\natexlab{}.
\newblock \showarticletitle{Toward 100TB Recommendation Models with Embedding Offloading}. In \bibinfo{booktitle}{\emph{Proceedings of the 18th ACM Conference on Recommender Systems}} (Bari, Italy) \emph{(\bibinfo{series}{RecSys '24})}. \bibinfo{publisher}{Association for Computing Machinery}, \bibinfo{address}{New York, NY, USA}, \bibinfo{pages}{841–843}.
\newblock
\showISBNx{9798400705052}
\href{https://doi.org/10.1145/3640457.3688037}{doi:\nolinkurl{10.1145/3640457.3688037}}


\bibitem[Paszke et~al\mbox{.}(2019)]%
        {NEURIPS2019_9015}
\bibfield{author}{\bibinfo{person}{Adam Paszke}, \bibinfo{person}{Sam Gross}, \bibinfo{person}{Francisco Massa}, \bibinfo{person}{Adam Lerer}, \bibinfo{person}{James Bradbury}, \bibinfo{person}{Gregory Chanan}, \bibinfo{person}{Trevor Killeen}, \bibinfo{person}{Zeming Lin}, \bibinfo{person}{Natalia Gimelshein}, \bibinfo{person}{Luca Antiga}, \bibinfo{person}{Alban Desmaison}, \bibinfo{person}{Andreas Kopf}, \bibinfo{person}{Edward Yang}, \bibinfo{person}{Zachary DeVito}, \bibinfo{person}{Martin Raison}, \bibinfo{person}{Alykhan Tejani}, \bibinfo{person}{Sasank Chilamkurthy}, \bibinfo{person}{Benoit Steiner}, \bibinfo{person}{Lu Fang}, \bibinfo{person}{Junjie Bai}, {and} \bibinfo{person}{Soumith Chintala}.} \bibinfo{year}{2019}\natexlab{}.
\newblock \showarticletitle{PyTorch: An Imperative Style, High-Performance Deep Learning Library}.
\newblock In \bibinfo{booktitle}{\emph{Advances in Neural Information Processing Systems 32}}. \bibinfo{publisher}{Curran Associates, Inc.}, \bibinfo{pages}{8024--8035}.
\newblock
\urldef\tempurl%
\url{http://papers.neurips.cc/paper/9015-pytorch-an-imperative-style-high-performance-deep-learning-library.pdf}
\showURL{%
\tempurl}


\bibitem[Raunak et~al\mbox{.}(2019)]%
        {raunak2019dim_reduction}
\bibfield{author}{\bibinfo{person}{Vikas Raunak}, \bibinfo{person}{Vivek Gupta}, {and} \bibinfo{person}{Florian Metze}.} \bibinfo{year}{2019}\natexlab{}.
\newblock \showarticletitle{Effective dimensionality reduction for word embeddings}. In \bibinfo{booktitle}{\emph{Proceedings of the 4th Workshop on Representation Learning for NLP (RepL4NLP-2019)}}. \bibinfo{pages}{235--243}.
\newblock


\bibitem[Reimers and Gurevych(2019)]%
        {reimers-2019-sentence-bert}
\bibfield{author}{\bibinfo{person}{Nils Reimers} {and} \bibinfo{person}{Iryna Gurevych}.} \bibinfo{year}{2019}\natexlab{}.
\newblock \showarticletitle{Sentence-BERT: Sentence Embeddings using Siamese BERT-Networks}. In \bibinfo{booktitle}{\emph{Proceedings of the 2019 Conference on Empirical Methods in Natural Language Processing}}. \bibinfo{publisher}{Association for Computational Linguistics}.
\newblock
\urldef\tempurl%
\url{http://arxiv.org/abs/1908.10084}
\showURL{%
\tempurl}


\bibitem[Templeton et~al\mbox{.}(2024)]%
        {templeton2024scaling}
\bibfield{author}{\bibinfo{person}{Adly Templeton}, \bibinfo{person}{Tom Conerly}, \bibinfo{person}{Jonathan Marcus}, \bibinfo{person}{Jack Lindsey}, \bibinfo{person}{Trenton Bricken}, \bibinfo{person}{Brian Chen}, \bibinfo{person}{Adam Pearce}, \bibinfo{person}{Craig Citro}, \bibinfo{person}{Emmanuel Ameisen}, \bibinfo{person}{Andy Jones}, {et~al\mbox{.}}} \bibinfo{year}{2024}\natexlab{}.
\newblock \bibinfo{title}{Scaling monosemanticity: Extracting interpretable features from claude 3 sonnet. Transformer Circuits Thread}.
\newblock


\bibitem[Virtanen et~al\mbox{.}(2020)]%
        {2020SciPy-NMeth}
\bibfield{author}{\bibinfo{person}{Pauli Virtanen}, \bibinfo{person}{Ralf Gommers}, \bibinfo{person}{Travis~E. Oliphant}, \bibinfo{person}{Matt Haberland}, \bibinfo{person}{Tyler Reddy}, \bibinfo{person}{David Cournapeau}, \bibinfo{person}{Evgeni Burovski}, \bibinfo{person}{Pearu Peterson}, \bibinfo{person}{Warren Weckesser}, \bibinfo{person}{Jonathan Bright}, \bibinfo{person}{St{\'e}fan~J. {van der Walt}}, \bibinfo{person}{Matthew Brett}, \bibinfo{person}{Joshua Wilson}, \bibinfo{person}{K.~Jarrod Millman}, \bibinfo{person}{Nikolay Mayorov}, \bibinfo{person}{Andrew R.~J. Nelson}, \bibinfo{person}{Eric Jones}, \bibinfo{person}{Robert Kern}, \bibinfo{person}{Eric Larson}, \bibinfo{person}{C~J Carey}, \bibinfo{person}{{\.I}lhan Polat}, \bibinfo{person}{Yu Feng}, \bibinfo{person}{Eric~W. Moore}, \bibinfo{person}{Jake {VanderPlas}}, \bibinfo{person}{Denis Laxalde}, \bibinfo{person}{Josef Perktold}, \bibinfo{person}{Robert Cimrman}, \bibinfo{person}{Ian Henriksen}, \bibinfo{person}{E.~A. Quintero},
  \bibinfo{person}{Charles~R. Harris}, \bibinfo{person}{Anne~M. Archibald}, \bibinfo{person}{Ant{\^o}nio~H. Ribeiro}, \bibinfo{person}{Fabian Pedregosa}, \bibinfo{person}{Paul {van Mulbregt}}, {and} \bibinfo{person}{{SciPy 1.0 Contributors}}.} \bibinfo{year}{2020}\natexlab{}.
\newblock \showarticletitle{{{SciPy} 1.0: Fundamental Algorithms for Scientific Computing in Python}}.
\newblock \bibinfo{journal}{\emph{Nature Methods}}  \bibinfo{volume}{17} (\bibinfo{year}{2020}), \bibinfo{pages}{261--272}.
\newblock
\href{https://doi.org/10.1038/s41592-019-0686-2}{doi:\nolinkurl{10.1038/s41592-019-0686-2}}


\bibitem[Wen et~al\mbox{.}(2025)]%
        {wen2025matryoshkarevisitingsparsecoding}
\bibfield{author}{\bibinfo{person}{Tiansheng Wen}, \bibinfo{person}{Yifei Wang}, \bibinfo{person}{Zequn Zeng}, \bibinfo{person}{Zhong Peng}, \bibinfo{person}{Yudi Su}, \bibinfo{person}{Xinyang Liu}, \bibinfo{person}{Bo Chen}, \bibinfo{person}{Hongwei Liu}, \bibinfo{person}{Stefanie Jegelka}, {and} \bibinfo{person}{Chenyu You}.} \bibinfo{year}{2025}\natexlab{}.
\newblock \bibinfo{title}{Beyond Matryoshka: Revisiting Sparse Coding for Adaptive Representation}.
\newblock
\showeprint[arxiv]{2503.01776}~[cs.LG]
\urldef\tempurl%
\url{https://arxiv.org/abs/2503.01776}
\showURL{%
\tempurl}


\end{thebibliography}

\end{document}